\documentclass{ceurart}

\usepackage{libertinus}
\usepackage{graphicx}
\usepackage[numbers]{natbib}
\usepackage{listings}
\begin{document}

\copyrightyear{2026}
\copyrightclause{Copyright for this paper by its authors.
  Use permitted under Creative Commons License Attribution 4.0
  International (CC BY 4.0).}

\conference{ISWC 2026}

\title{Ask the Curator: Demonstrating Expert-Driven RDF Data Curation with HERITRACE}

\author[1]{Arcangelo Massari}[%
orcid=0000-0002-8420-0696,
email=arcangelo.massari@unibo.it]
\author[1]{Silvio Peroni}[%
orcid=0000-0003-0530-4305,
email=silvio.peroni@unibo.it]
\address[1]{Digital Humanities Advanced Research Centre (/DH.arc), Department of Classical Philology and Italian Studies, Alma Mater Studiorum -- Universit\`a di Bologna, Italy}

\begin{abstract}
HERITRACE is an open-source Web application for curating RDF data held in triplestores, with provenance and change tracking
recorded in RDF as well. It is agnostic to the data model: a technician
configures it for a collection by writing SHACL shapes and YAML display rules,
after which domain experts edit the data through generated forms, and every
change becomes a provenance snapshot that can be inspected and restored. This
paper traces the path from configuration to curation on a documented
case of bibliographic ambiguity: two unrelated articles that PubMed records
under the same DOI. We show how a technician prepares a small OpenCitations
Meta subset for curation and how a curator acts on duplicate suggestions to merge
two articles, inadvertently confirms a further merge despite conflicting
metadata, reverts the mistake through the Time Machine, and corrects the DOI
after verification against Crossref.
\end{abstract}


\maketitle

\section{Introduction}

Curating RDF data requires two competences that rarely coincide in one
person: familiarity with the domain the data describe, and fluency in Semantic Web technologies, particularly RDF, SPARQL,
OWL ontologies, and the machinery of triplestores.
HERITRACE~\cite{massari2026heritrace} separates them. It is a Web application
that connects to existing SPARQL-queryable stores without data migration and
derives its editing interface from declarative configuration: SHACL
shapes~\cite{shacl2017} define the data model and its constraints, while YAML
display rules configure aspects beyond that scope, such as presentation and the
criteria used to identify potential duplicates. Domain experts work only with
the resulting forms. Each creation,
modification, merge, and deletion is recorded as a provenance snapshot in RDF
following the OpenCitations Data Model
(OCDM)~\cite{daquino2020ocdm}, so the change history is as findable, accessible, interoperable, and
reusable~\cite{wilkinson2016fair} as the curated resources, queryable and
restorable with the same technologies. The architecture and
evaluation of the system are presented in a separate
paper~\cite{massari2026heritrace}; here we demonstrate its use in practice, from initial
configuration to the correction of a real case of bibliographic ambiguity.

The walkthrough is organised around a decision that resists automation. It
is grounded in a real case from OpenCitations Meta~\cite{massari2024meta}, an
aggregator of scholarly bibliographic metadata that collects records from
heterogeneous sources, including Crossref~\cite{hendricks2020crossref} and
PubMed~\cite{sayers2026ncbi}. To deduplicate the ingested records,
OpenCitations Meta relies on shared identifiers rather than heuristics,
favouring precision at the expense of recall. Even identifiers, however, can
carry errors inherited from the original sources, producing collisions that
are difficult to resolve in a fully automatic and general way. The demonstration dataset provided in this paper reproduces one such case with real records:
PubMed assigns the same DOI to two unrelated articles of the
\textit{Scandinavian Journal of Infectious Diseases}, while Crossref assigns
it to only one of them. By design, duplicate detection in HERITRACE only suggests candidates. The curator reviews the source records and decides which are genuine duplicates before acting.

The rest of the paper is organised as follows. Section~\ref{sec:related} briefly mentions some related systems. Section~\ref{sec:data} introduces the demonstration
data. Section~\ref{sec:setup} presents the technician's configuration, while 
Section~\ref{sec:walkthrough} introduces the curator's session. A
video of the session\footnote{\url{https://youtu.be/EuFtxQoJvPg}, also archived on Zenodo~\cite{massari2026demo_package}} and a reproducible package~\cite{massari2026demo_package} accompany the paper. Finally, Section~\ref{sec:conclusions} concludes the paper, sketching out some future work.

\section{Related Works}
\label{sec:related}

Several systems provide editing interfaces for RDF data; for instance, Wikibase~\cite{diefenbachWikibaseInfrastructureKnowledge2021}, ResearchSpace~\cite{oldmanReshapingKnowledgeGraph2018}, and CLEF~\cite{daquinoCLEFLinkedOpen2023,giacominiCLEF20Solutions2025}. Sch{\'i}matos~\cite{10.1007/978-3-030-62466-8_5} shares with HERITRACE the approach of generating Web forms from SHACL shapes. The companion paper~\cite{massari2026heritrace} presents a detailed functional comparison with these systems.

\section{Demonstration Data}
\label{sec:data}

The scenario involves two articles published in the \textit{Scandinavian
Journal of Infectious Diseases}: one by Nilsson \textit{et
al.}~\cite{nilsson2005fourth} (PMID \texttt{15849057}) and one by Maltezou
\textit{et al.}~\cite{maltezou2004mycoplasma} (PMID \texttt{15370649}).
Crossref\footnote{\url{https://api.crossref.org/works/10.1080\%2F00365540410020884}}
assigns the DOI \texttt{10.1080/00365540410020884} to the Nilsson article
alone, but PubMed lists it in both
records\footnote{\url{https://eutils.ncbi.nlm.nih.gov/entrez/eutils/esummary.fcgi?db=pubmed&id=15370649,15849057&retmode=json}}.

The demonstration graph mirrors what OpenCitations Meta would obtain by
importing both PubMed records and the Crossref record. It is modelled as a
small subset of the live collection, which describes bibliographic entities
with the OCDM:
articles are \texttt{fabio:JournalArticle} resources, and each identifier is
a \texttt{datacite:Identifier} node linked through the
\texttt{datacite:hasIdentifier} property. The graph contains three RDF descriptions of
the two articles, hereafter labelled \textbf{NC} (the Nilsson article imported
from Crossref), \textbf{NP} (the Nilsson article imported from PubMed), and
\textbf{MP} (the Maltezou article imported from PubMed). NC, NP, and MP all
initially reference one shared \texttt{datacite:Identifier} node whose literal
value is \texttt{10.1080/00365540410020884}. NC and NP
agree on title and publication details. MP differs in
title, authors, date, volume, issue, and pages; its only misleading identity
signal is the shared DOI node.

Curation should end with NP merged into NC, carrying the DOI and PMID
\texttt{15849057}, and MP carrying only PMID \texttt{15370649}
with no link to the shared DOI node.

\section{Preparing the Curation Environment}
\label{sec:setup}

The technician prepares three configuration layers: the deployment
environment, the data model constraints, and the display rules.

Listing~\ref{lst:compose} shows the Docker Compose configuration excerpt.
HERITRACE runs alongside two triplestores---one for the dataset, one for its
provenance---and connects to them through their SPARQL endpoints
(\texttt{DATASET\_DB\_URL}, \texttt{PROVENANCE\_DB\_URL}).
External datasets that adopt HERITRACE as their curation system may not follow
the OpenCitations Data Model for provenance and change tracking, or may lack
provenance data entirely. To handle this, HERITRACE allows the
deployer to declare a \texttt{PRIMARY\_SOURCE} and a
\texttt{DATASET\_GENERATION\_TIME}: in the demonstration these point to the
OpenCitations Meta dump from which the PubMed records were first
ingested~\cite{opencitations2023meta_dump_v4} and its release time. When an entity is first
modified, HERITRACE lazily generates its initial provenance snapshot, recording
these values as \texttt{prov:hadPrimarySource} and
\texttt{prov:wasGeneratedAtTime}. The two domain-specific files---a SHACL
graph and a YAML rule set---are mounted read-only into the application
container.

\begin{lstlisting}[caption={Docker Compose configuration (excerpt)},label=lst:compose]
services:
  dataset-db:
    image: arcangelo7/heritrace-iswc2026-database:1.0.0
    environment:
      CONTAINER_TYPE: dataset
  provenance-db:
    image: arcangelo7/heritrace-iswc2026-database:1.0.0
    environment:
      CONTAINER_TYPE: provenance
  web:
    image: heritrace:iswc2026-dev
    environment:
      DATASET_DB_URL: http://dataset-db:8890/sparql
      PROVENANCE_DB_URL: http://provenance-db:8890/sparql
      DATASET_GENERATION_TIME: "2023-06-28T13:00:48+00:00"
      PRIMARY_SOURCE: https://doi.org/10.6084/m9.figshare.21747536.v4
      BASE_IRI: https://w3id.org/oc/meta/demo
      # ...
    volumes:
      - ./config/shacl.ttl:/app/shacl.ttl:ro
      - ./config/display_rules.yaml:/app/display_rules.yaml:ro
\end{lstlisting}

Listing~\ref{lst:shacl} shows the SHACL shapes that define the journal
article entity type. The identifier shape constrains each node to carry
exactly one scheme---\texttt{datacite:doi} or
\texttt{datacite:pmid}---and one literal value. The article shape references
it and further requires a title, publication date, authors, venue, and pages.
From these shapes HERITRACE builds the editing forms and validates curator
input.

\begin{lstlisting}[caption={SHACL shapes (excerpt)},label=lst:shacl]
schema:JournalArticleIdentifierShape a sh:NodeShape ;
  sh:targetClass datacite:Identifier ;
  sh:property [
    sh:path datacite:usesIdentifierScheme ;
    sh:in (datacite:doi datacite:pmid) ;
    sh:minCount 1 ; sh:maxCount 1 ] ;
  sh:property [
    sh:path literal:hasLiteralValue ;
    sh:datatype xsd:string ;
    sh:minCount 1 ; sh:maxCount 1 ] .

schema:JournalArticleShape a sh:NodeShape ;
  sh:targetClass fabio:JournalArticle ;
  sh:property [
    sh:path datacite:hasIdentifier ;
    sh:node schema:JournalArticleIdentifierShape ;
    sh:minCount 1 ] ;
  sh:property [
    sh:path dcterms:title ;
    sh:datatype xsd:string ;
    sh:minCount 1 ; sh:maxCount 1 ] ;
  # ... publicationDate, authors, venue, pages
  .
\end{lstlisting}

Listing~\ref{lst:display} shows the display rule for journal articles. The
\texttt{displayName} field assigns the human-readable label shown in the
interface, and each \texttt{displayProperties} entry configures how a property
appears in the editing form: \texttt{fetchValueFromQuery} references a SPARQL
query that resolves complex values such as reified identifiers into
human-readable strings and \texttt{inputType} selects the widget. The
\texttt{similarity\_properties} list configures duplicate detection with OR
semantics: an entity is suggested as a candidate when at least one configured
property matches. Each property is compared as a direct RDF value by exact
equality. For \texttt{datacite:hasIdentifier} the direct value is the
identifier node itself, which is why a shared identifier node is enough to
make two articles candidates for merging. Suggestions are restricted to
resources with the same configured RDF type.

\begin{lstlisting}[caption={Display rules for journal articles (excerpt)},label=lst:display]
- target:
    class: "http://purl.org/spar/fabio/JournalArticle"
  displayName: "Journal Article"
  similarity_properties:
    - "http://purl.org/dc/terms/title"
    - "http://purl.org/spar/datacite/hasIdentifier"
  displayProperties:
    - property: "http://purl.org/spar/datacite/hasIdentifier"
      displayName: "Identifier"
      fetchValueFromQuery: *identifier_query
    - property: "http://purl.org/dc/terms/title"
      displayName: "Title"
      inputType: "textarea"
    # ... authors, date, venue, pages
\end{lstlisting}

\section{Curation Walkthrough}
\label{sec:walkthrough}

\begin{figure}
  \centering
  \begin{minipage}[t]{0.49\linewidth}
    \centering
    \includegraphics[width=\linewidth]{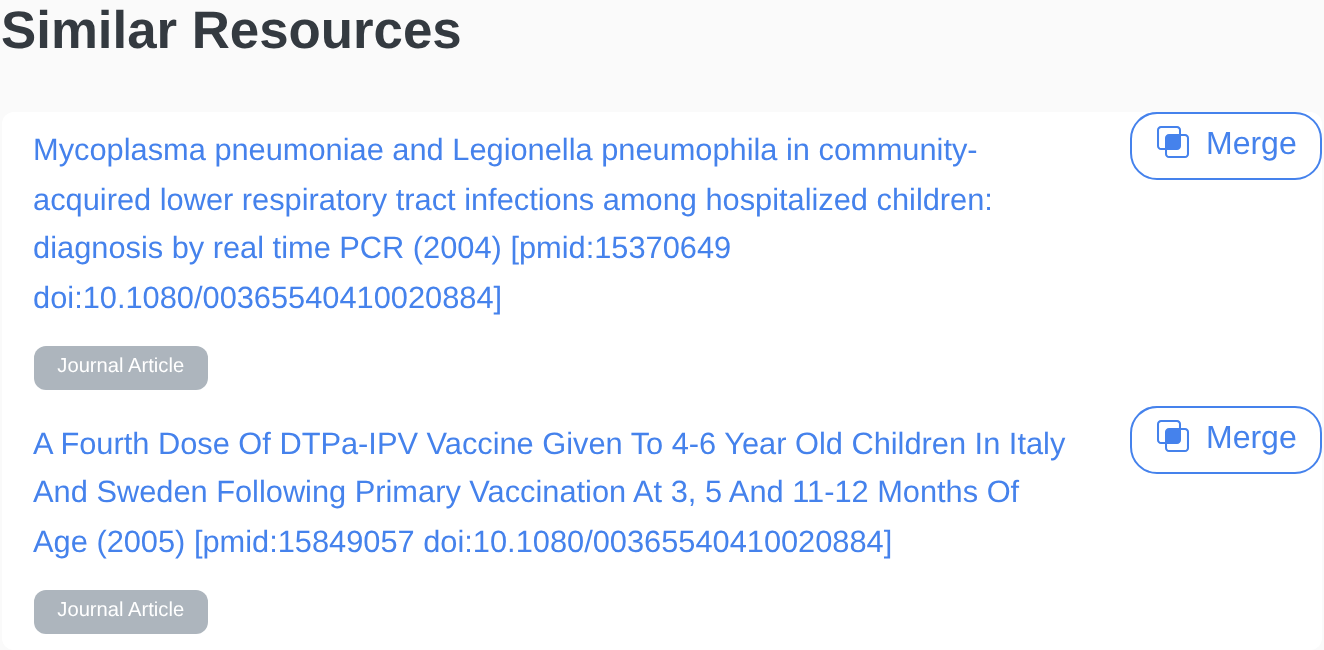}
    \par\smallskip\textbf{(a)}
  \end{minipage}\hfill
  \begin{minipage}[t]{0.49\linewidth}
    \centering
    \includegraphics[width=\linewidth,trim=0 1720bp 0 0,clip]{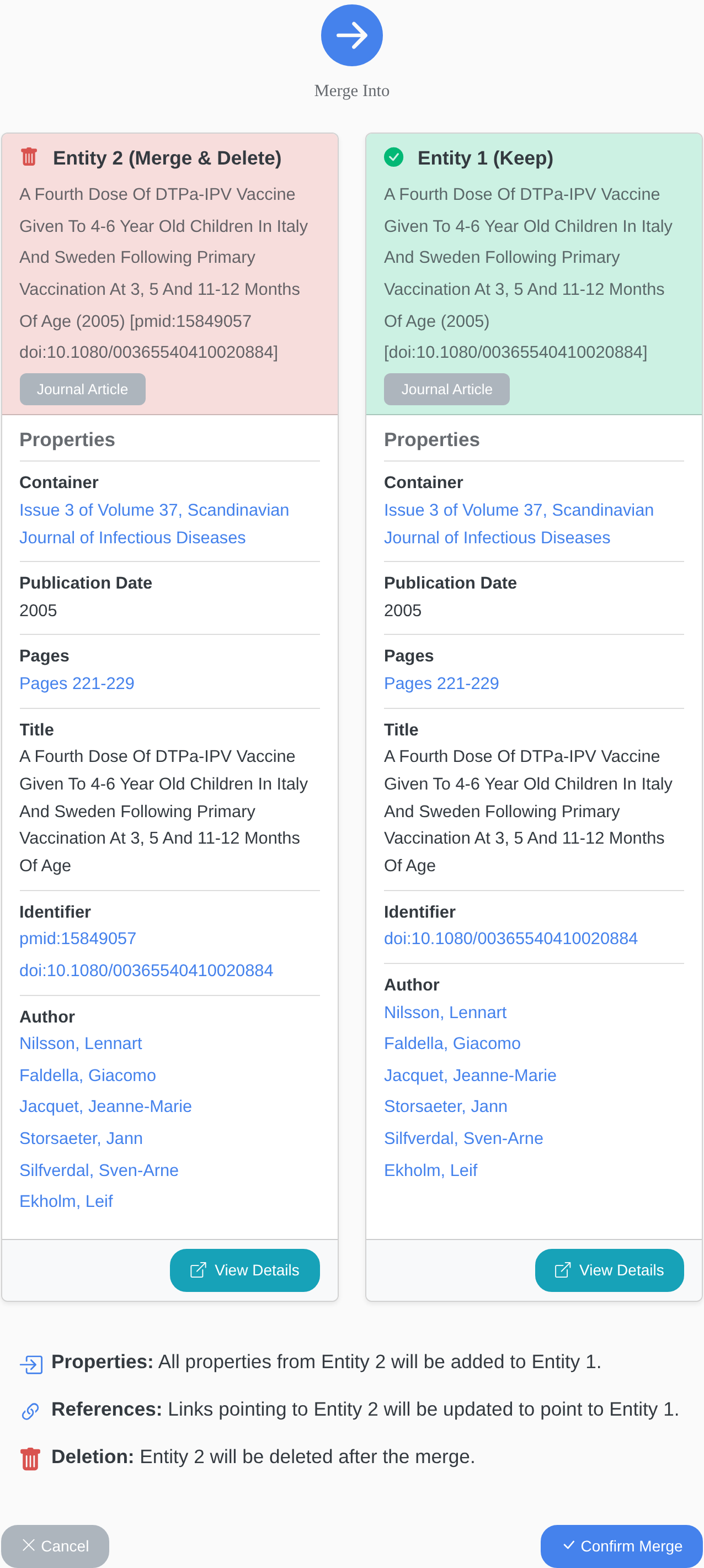}
    \par\smallskip\textbf{(b)}
  \end{minipage}
  \caption{HERITRACE presents (a) two duplicate candidates for NC
    and (b) the comparison with NP.}
  \label{fig:merge}
\end{figure}

The curator opens NC. HERITRACE
lists two merge candidates (Figure~\ref{fig:merge}(a)): NP, which matches on
both configured properties
(its title equals that of NC, and it references the same
identifier node), and MP, which matches only through the
shared DOI node.

A merge begins with the entity to retain; the interface presents a
side-by-side comparison of the two descriptions (Figure~\ref{fig:merge}(b)),
and on confirmation transfers statements and incoming links from the merged
entity to the retained one and deletes the merged entity. The interface
prompts the curator to select a primary source for the operation, recorded
in the provenance graph as \texttt{prov:hadPrimarySource}; the curator
indicates the PubMed API response for PMID \texttt{15849057}, identifying the
origin of the absorbed description. NC now also carries that PMID.

The remaining suggestion is more deceptive: it is produced by the shared DOI
node, yet the comparison shows that title, authors, date, volume, issue, and
pages all refer to another article. Despite these visible conflicts, the
curator confirms the merge by mistake, selecting the PubMed API
response for PMID \texttt{15370649} as its primary source. The graph now conflates the Nilsson and
Maltezou works.

\begin{figure}
  \centering
  \begin{minipage}[b]{0.44\linewidth}
    \centering
    \includegraphics[width=\linewidth]{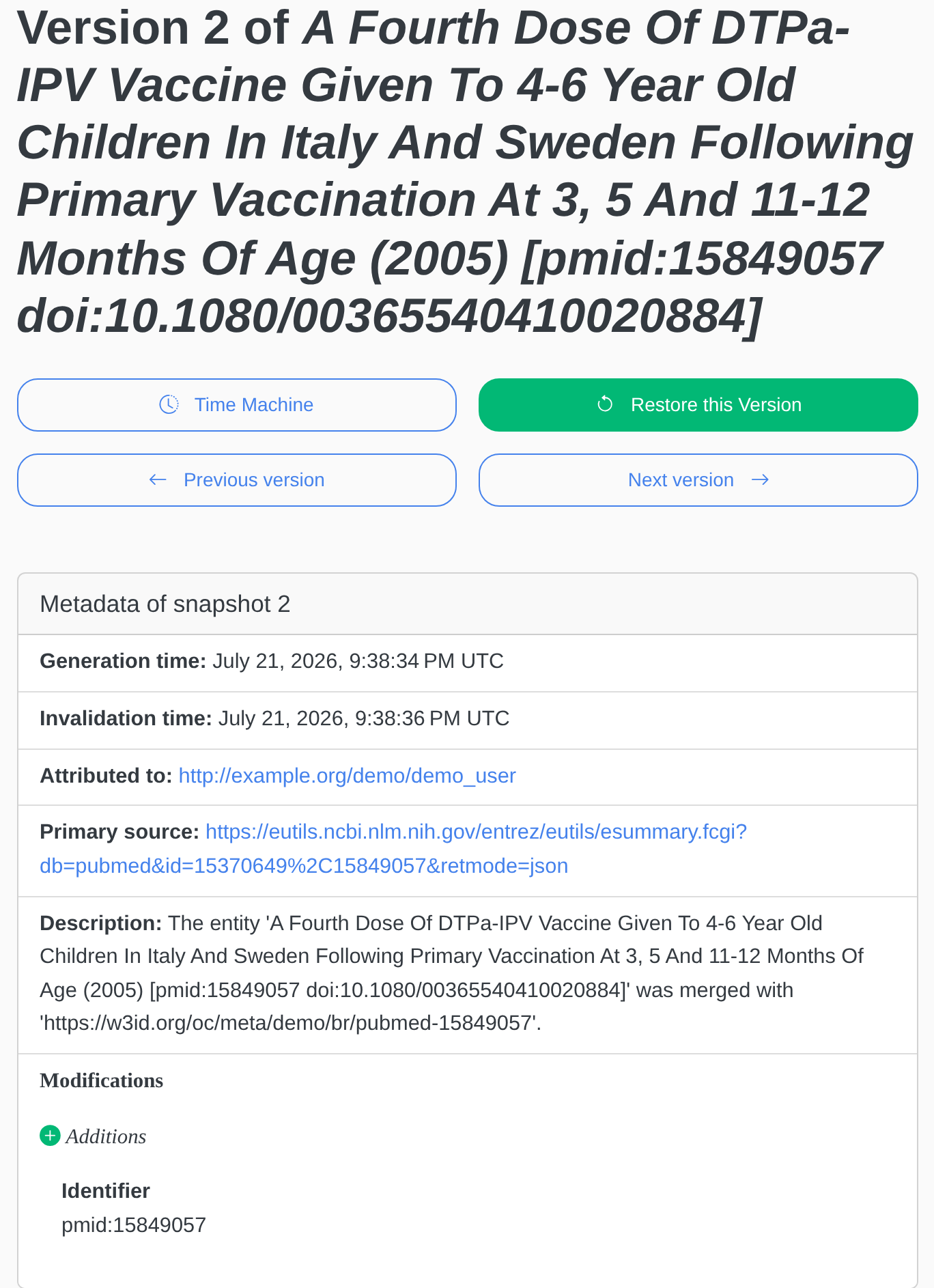}
    \par\smallskip\textbf{(a)}
  \end{minipage}\hfill
  \begin{minipage}[b]{0.54\linewidth}
    \centering
    \includegraphics[width=\linewidth]{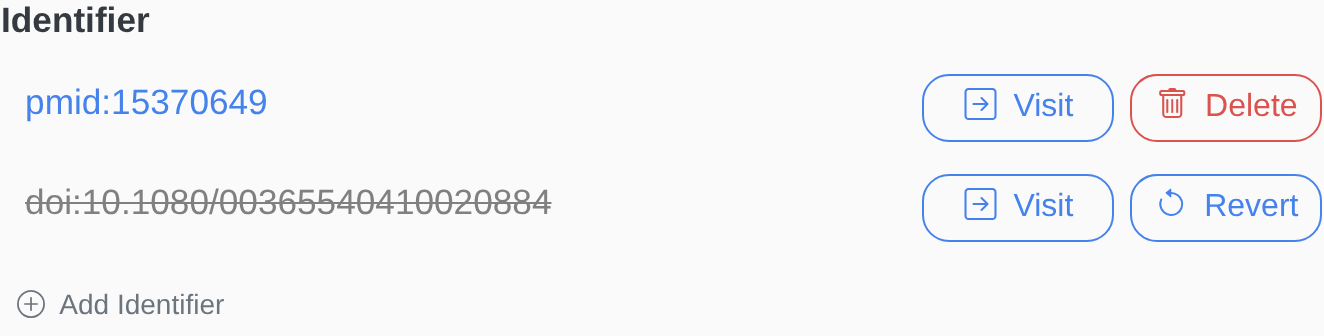}
    \par\smallskip\textbf{(b)}
    \par\medskip
    \includegraphics[width=\linewidth]{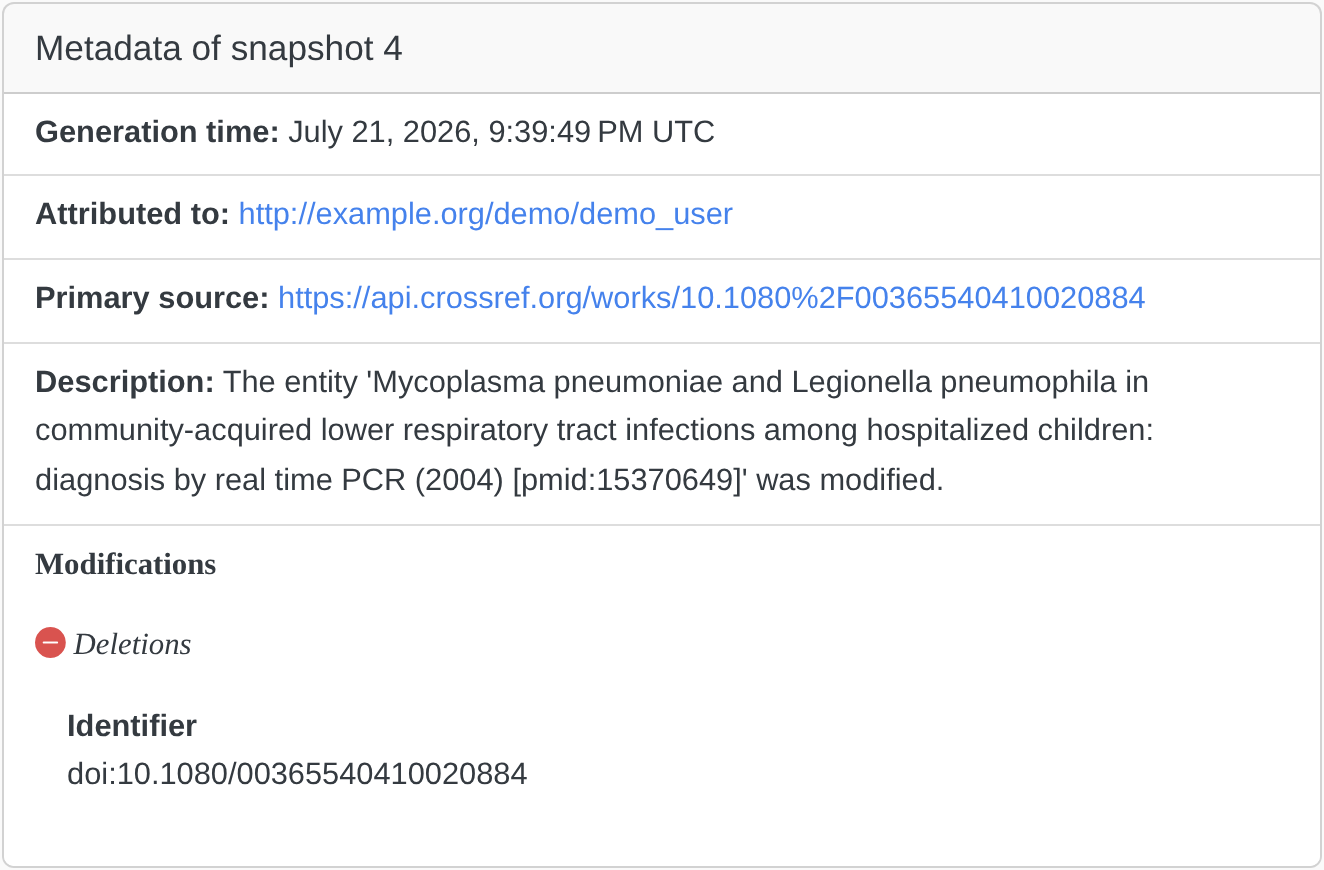}
    \par\smallskip\textbf{(c)}
  \end{minipage}
  \caption{The curator (a) selects the version preceding the MP merge,
    (b) marks the shared DOI for removal while retaining the PMID, and (c)
    inspects the correction snapshot and its Crossref API provenance.}
  \label{fig:history}
\end{figure}

Because both decisions have become provenance snapshots, the curator can
reconstruct how the conflation occurred. The Time Machine presents each
version with its timestamp, responsible agent, primary source, and change
description. Scrolling through the timeline, the curator locates the snapshot
that immediately follows the correct NC--NP merge and precedes the accidental
MP merge (Figure~\ref{fig:history}(a)), then selects \textit{Restore this
Version}. This rolls NC back to that intermediate state, reversing only the
accidental merge: NC and NP remain combined, whereas MP reappears as a
separate entity.

Restoration removes the conflation but leaves the cause of the false
suggestion in place, since both entities still reference the same DOI node.
The curator therefore checks the DOI through the Crossref API response, which
assigns it to the Nilsson article. On this evidence, the curator opens
MP and unlinks the shared DOI node
(Figure~\ref{fig:history}(b)), leaving PMID \texttt{15370649} as its only
identifier. HERITRACE records the correction and the Crossref response used as
its primary source (Figure~\ref{fig:history}(c)). The sequence of snapshots
now connects each decision to its evidence, and the graph correctly represents
two distinct works: NC with DOI and PMID, and MP with its
PMID alone.

\enlargethispage{4\baselineskip}
\section{Conclusions and Future Works}
\label{sec:conclusions}

Several directions remain open. Both the usability study reported in the
companion paper~\cite{massari2026heritrace} and the walkthrough presented here
involved bibliographic metadata experts; a cross-domain evaluation would test
whether the domain-independent interaction model generalises to other fields.
The walkthrough also assumed a single curator, whereas production curation is
typically collaborative: extending the system to handle concurrent
modifications would require conflict resolution mechanisms, building on the
per-change attribution the provenance model already provides. Duplicate
detection currently relies on exact property matching; complementing it with
similarity-based comparisons would surface candidates that describe the same
entity without sharing any identifier. Finally, the file-based configuration requires a technician for every data
model change. Because HERITRACE already generates editing forms from RDF
descriptions, expressing the display rules in RDF as well, for instance through
YAML-LD~\cite{kelloggYAMLLD2026}, would allow it to generate forms for editing
its own SHACL shapes~\cite{shacl2017} and display rules, making the system a
configuration interface for itself.

\paragraph*{Acknowledgments.}
This work was funded by the European Union's Horizon Europe programme, Grant Agreement No.\@ 101188018 (GRAPHIA).

\section*{Declaration on Generative AI}
The authors have not employed any Generative AI tools.

\bibliography{main}

\end{document}